\documentclass[a4paper,11pt]{article}
\usepackage{jcappub} 

\usepackage{amsmath,amssymb,bbold,epsfig}
\usepackage{amsfonts}
\usepackage{slashed}
\usepackage{dsfont}
\usepackage{graphicx,color}
\usepackage{bbm}
\usepackage{bm}
\usepackage[normalem]{ulem}

\vspace*{-1.5cm}
\title{\mbox{Relic neutrino detection through angular} correlations 
in inverse $\beta$-decay}

\author{Evgeny~Akhmedov}

\affiliation{
Max-Planck-Institut f\"ur Kernphysik, Saupfercheckweg 1, \\69117 Heidelberg, 
Germany
}
 
\emailAdd{akhmedov@mpi-hd.mpg.de}

\abstract{
Neutrino capture on beta-decaying nuclei is currently the only known 
potentially viable method of detection of cosmic background neutrinos.  
It is based on the idea of separation of the spectra of electrons 
or positrons produced in captures of relic neutrinos on 
unstable nuclei from those from the usual $\beta$-decay and requires 
very high energy resolution of the detector, comparable to the neutrino 
mass. In this paper we suggest an alternative method of discrimination 
between neutrino capture and $\beta$-decay, based on periodic variations 
of angular correlations in inverse beta decay transitions induced by relic 
neutrino capture. The time variations are expected to arise due to 
the peculiar motion of the Sun with respect to the C$\nu$B rest frame and 
the rotation of the Earth about its axis and can be observed in experiments 
with both polarized and unpolarized nuclear targets. The main advantage of 
the suggested method is that it does not depend crucially on the energy 
resolution of detection of the produced $\beta$-particles and can be operative 
even if this resolution exceeds the largest neutrino mass. 
}

\begin{document}
\maketitle
\flushbottom

\newcommand{\be}{\begin{equation}}
\newcommand{\ee}{\end{equation}}
\newcommand{\bea}{\begin{eqnarray}}
\newcommand{\eea}{\end{eqnarray}}

\section{\label{sec:intro}Introduction}
Standard cosmology predicts the existence of a sea of active left-handed 
neutrinos and their right-handed antiparticles that decoupled from the cosmic 
plasma at temperatures $T\sim 2$ MeV and have cooled down in the course of 
the expansion of the Universe \cite{dolgov1,lesg1,quigg}. At present, these 
cosmic background neutrinos (C$\nu$B) are expected to have nearly Fermi-Dirac 
spectrum with the temperature $T_{\nu 0}\simeq 1.945$\,K $\simeq 1.676\times 
10^{-4}$ eV and to be in states of definite mass rather than in flavour 
states \cite{tom1}. Together with data from neutrino oscillation experiments, 
this, in particular, means that at least two relic neutrino species must be 
non-relativistic at the present epoch. Because of their extremely low energies 
and the weak nature of neutrino interactions, the relic neutrinos have not yet 
been directly detected. At the same time, they 
should 
carry rich and important information about the very early stages of the 
evolution of the Universe; their observation thus represents one of the main 
challenges of modern cosmology. Detection of relic neutrinos may also shed 
light on some important neutrino properties, such as their Dirac vs.\ Majorana 
nature and possible existence of light sterile neutrinos. 
 
Several approaches to detecting the C$\nu$B neutrinos have been suggested 
to date, see e.g.\ refs.~\cite{rev1,rev2,rev3} for reviews. Unfortunately, 
most of them either were based on flawed considerations or are rather 
impractical. Out of all the suggestions put forward so far, only neutrino 
capture on beta-decaying nuclei has a chance to bear fruit in a foreseeable 
future. As this process is threshold-free, even neutrinos of extremely low 
energies can be captured with finite rate. The approach is based on 
separation of the spectra of $\beta$-particles produced in capture of relic 
neutrinos from those coming from the usual $\beta$-decay 
of the target nuclei. It was first suggested by Weinberg back in 1962 in the 
context of massless relic neutrinos with large chemical potential $\xi$ 
\cite{weinb}. However, current stringent limits on $\xi$ rule out this 
possibility.  

The idea of using neutrino capture on beta-decaying nuclei for detecting 
C$\nu$B neutrinos was revived in a modern setting by Cocco et al. 
\cite{cocco1}, who noted that, as the neutrinos are actually massive, 
the spectrum of $\beta$-particles 
produced in captures of relic neutrinos of mass $m_\nu$  
must be separated from the spectrum of the much more abundant electrons or 
positrons coming from the usual $\beta$ decay of the target nuclei by a gap 
$\sim 2m_\nu$ (assuming $m_\nu\gtrsim T_{\nu 0}$). Then, if the 
$\beta$-particles are detected with the energy resolution better than 
$m_\nu$, one can tell these two processes apart. Following this observation, 
various aspects of relic neutrino detection through their capture on 
beta-decaying nuclei were studied in a number of publications, see e.g. 
\cite{vogel1,blennow,cocco2,li,lusignoli,faessler,safdi,long,lisanti1,roulet,
lee}. 

The necessity  of separation of the spectra of $\beta$-particles 
produced in relic neutrino capture from those coming from the usual $\beta$ 
decay of the target nuclei puts extremely challenging demands 
on the energy resolution of the detection. 
Barring the existence of a light predominantly sterile neutrino with a 
sizeable mixing to the electron flavour and significant presence in 
C$\nu$B and assuming that the neutrino mass spectrum is hierarchical rather 
than quasi-degenerate, one finds that the energy resolution 
of at least 0.05 eV is necessary. At present, there is one experimental 
proposal for relic neutrino detection through their capture on beta-decaying 
nuclei -- the PTOLEMY experiment \cite{ptol1,ptol2,ptol3}, which plans to use 
tritium as the target. The collaboration has set 
an ambitious goal of achieving the energy resolution of 0.05 eV for electron 
detection. 
It should be noted, however, that in the case of normal neutrino mass ordering 
(which is currently preferred by the neutrino oscillation data at $3\sigma$ 
level \cite{hierarchy1,hierarchy2,hierarchy3}) and very small mass of the 
lightest neutrino, even an order of magnitude smaller 
energy resolution may be necessary \cite{blennow,long}. This is related to 
the fact that in this case the heaviest neutrino with the mass 
$m_3\simeq 0.05$ eV 
has only a small electron neutrino component $|U_{e3}|^2\simeq 2.2\times 
10^{-2}$. Alternatively, one would need a high enough statistics of the relic 
neutrino capture events in order to compensate for the smallness of 
$|U_{e3}|^2$. 

In the present paper we suggest an alternative method of detecting C$\nu$B 
neutrinos, based on observing {\sl angular correlations} that are 
characteristic of beta processes. While still exploiting neutrino capture on 
beta-decaying nuclei, it does not require discrimination between the 
$\beta$-particles coming from relic neutrino capture and from the usual 
$\beta$ decay by separating their spectra. Instead, we suggest to employ time 
variations of the capture rate of relic neutrinos on polarized or unpolarized 
nuclei arising due to the peculiar motion of the Sun and the rotation of the 
Earth about its axis. 

Neutrino capture on polarized nuclei exhibits several angular correlations, 
and in particular a correlation between the velocity of the incoming 
neutrinos and nuclear polarization. Due to the peculiar motion of the solar 
system with respect to the C$\nu$B rest frame, relic neutrinos 
have a preferred direction of arrival at the Earth. 
At the same time, if the direction of polarization of the target nuclei 
is fixed in the Earth-bound lab frame, the angle between this direction and 
that of the preferred velocity of relic neutrinos 
varies during the day due to the Earth's rotation about its axis. One can 
therefore expect time variation of the 
relic neutrino signal with the period equal to the sidereal day 
$T_0\simeq\;$23h 56m 4s.%
\footnote{The sidereal day is the period of the Earth's rotation about its 
axis with respect to the distant stars, which is slightly shorter than the 
solar day because of the Earth's orbital motion.
} 
No such time variation is expected for the rate of the $\beta$ decay of the 
target nuclei.

Relic neutrino capture on polarized nuclei, including possible time variations 
of the signal due to the peculiar motion of the Sun and the rotation 
of the Earth, has previously been considered by Lisanti et al. 
\cite{lisanti1} with application to tritium target. However, the authors of 
\cite{lisanti1} discussed this process essentially as a method of obtaining 
additional information about 
the properties of C$\nu$B neutrinos (such as anisotropies of their velocities 
and spin distributions) within the usual approach to relic neutrino detection 
based on the separation of spectra of electrons from neutrino capture and 
$\beta$ decay. By contrast, we focus on using various angular correlations 
in neutrino capture on polarized nuclear targets as a means of separation of 
the relic neutrino signal from the background.

We also consider the possibility of using angular correlations 
for detecting relic neutrinos with unpolarized nuclear targets. This 
can in principle be done either by measuring polarizations of final-state 
nuclei or by observing asymmetry of the produced $\beta$ particles with 
respect to the preferred direction of relic neutrino arrival at the Earth 
($\beta-\nu$ correlation). The latter should also give rise to a 
time-dependent signal if the $\beta$-asymmetry is studied with respect to a 
direction that is fixed in the Earth-bound lab frame. 

The amplitude of the time variation of the relic neutrino signal is expected 
to be small, and the problem of reliably detecting this variation in the 
presence of a large background is very challenging. The main difficulty comes 
actually not from the high average background level but rather from the 
fluctuations of the background. Fortunately, there are well-developed methods 
of such signal-from-noise separation, which are especially efficient when 
the signal has a known periodicity. 

The main advantage of the proposed approach is that it does not 
depend crucially on the energy resolution 
of detection of the produced $\beta$-particles. While good energy resolution 
would help to suppress the background from the usual $\beta$ decay by allowing 
one to work close to the endpoint of the $\beta$-spectrum, the method can 
still work even if the resolution is relatively large, provided that a 
sufficiently powerful method of separation of the weak periodic signal from 
strong random noise is employed. At the same time, the approach based on 
separation of the spectra will be merely inoperative 
if the energy resolution of the detection exceeds the largest neutrino mass.  
One consequence of the fact that the requirements on the energy resolution 
are less severe in the approach we suggest is that radioactive nuclei with 
larger  $Q_\beta$ values may be preferable as target, as they lead to larger 
absolute detection rates. 

In our discussion of relic neutrino capture on 
polarized nuclei we concentrate on pure Gamow-Teller transitions, 
taking the allowed $1^\pi\to 0^\pi$ nuclear transitions as an example. 
Our results can, however, be readily extended to other Gamow-Teller 
transitions. One advantage of pure Gamow-Teller transitions is that 
the effects of target polarization 
are in general more pronounced in this case. On the other hand, pure Fermi 
transitions $0^\pi\to 0^\pi$ (as well as pure Gamow-Teller ones) may be useful 
when considering $\beta-\nu$ angular correlations in relic neutrino capture 
on unpolarized nuclei. For mixed transitions the correlation coefficient may 
be strongly suppressed. Another important advantage of pure transitions is 
that for such transitions the angular correlation coefficients 
do not depend on nuclear 
matrix elements. 

The paper is organized as follows. In section~\ref{sec:GT} we derive the 
expression for the differential cross section of neutrino capture on polarized 
nuclei in the case of allowed Gamow-Teller $1^\pi\to 0^\pi$ transitions and 
discuss various angular correlations relevant to this process. 
In section~\ref{sec:betanu} we consider angular correlations in the case of 
neutrino capture on unpolarized nuclei. 
In section~\ref{sec:trit} we compare the results of section~\ref{sec:GT} 
with those obtained for neutrino capture on polarized tritium~\cite{lisanti1}. 
In section~\ref{sec:movframe} we consider effects of averaging over the 
directions of relic neutrinos on angular correlations in neutrino capture 
on polarized and unpolarized nuclei observed in the lab frame.
In section~\ref{sec:signoise} we discuss the problem of extraction of 
weak periodic signals from large fluctuating backgrounds and its application 
to relic neutrino detection. 
Our results are summarized and discussed in section~\ref{sec:disc}. 
Appendix A contains some technical details related to the averaging over 
angular distributions of relic neutrinos. Electron angular asymmetries with 
respect to fixed directions in the lab frame in experiments with unpolarized 
targets are considered in Appendix B.

\section{\label{sec:capt}
Neutrino capture in inverse $\beta$ decay}

We shall consider neutrino detection in the inverse 
$\beta^-$-decay process   
\be
\nu_j(q)+A_i(p)\to A_f(p')+e^-(k)\,, 
\label{eq:proc1}
\ee
where $\nu_j$ is the $j$th neutrino mass eigenstate, $A_i$ and $A_f$ are the 
parent and daughter nuclei, and the 4-momenta of the participating particles 
are indicated in the parentheses. Detection of relic (anti)neutrino states in 
inverse $\beta^+$ decays can be considered quite similarly. 

At present, C$\nu$B neutrinos originally produced in the states of left-handed 
chirality should be in left-handed helicity eigenstates in the C$\nu$B rest 
frame, and similarly for the states of right-handed chirality and helicity 
\cite{long}. Note that, in the Dirac neutrino case, only neutrinos (which are 
left-helical now) can be captured in the $\beta^-$-process (\ref{eq:proc1}), 
while right-helical antineutrinos can be detected in inverse 
$\beta^+$-processes. At the same time, if neutrinos are Majorana particles, 
for non-relativistic neutrinos both left-helical and right-helical states can 
participate in process (\ref{eq:proc1}) through their left-chirality 
components \cite{long}.  

As helicity is not a Lorentz-invariant quantity, neutrinos that are in helicity 
eigenstates in some frame $K$ do not in general have definite helicity in a 
frame $K'$ moving with respect to $K$. Because the Earth moves 
with respect to the C$\nu$B rest frame, the spins 
of relic neutrinos need not be aligned or antialigned with their velocities 
in the lab frame. We therefore consider the differential cross sections of 
neutrino detection for arbitrary directions of the spins of the initial-state 
neutrinos. 

\subsection{\label{sec:GT}Gamow-Teller $1^\pi\to 0^\pi$ transitions}

For definiteness, we focus on the allowed $1^\pi\to 0^\pi$ Gamow-Teller 
nuclear transitions, some examples being the decays $^{32}$P$\to^{32}$S, 
$^{64}$Co$\to^{64}$Ni and $^{80}$Br$\to^{80}$Kr. We will initially consider 
the parent nucleus, the incoming neutrino and the produced electron to be in 
definite spin states, that is, no summation or averaging over their spins is 
performed. The differential cross section of the process $d\sigma_j$ 
multiplied by the neutrino velocity $v_j$ can then be found \vspace*{1.5mm} 
as  
\be
v_j d\sigma_j=\frac{G_\beta^2}{2}
|U_{ej}|^2
\frac{1}{(2\pi)^2}
\frac{(\varepsilon_{\mu}\varepsilon_{\nu}^* X^{\mu\nu})}
{4E_e E_j}
|M_{\rm GT}|^2 
F(Z, E_e) 
E_e \sqrt{E_e^2-m_e^2}\,
d\Omega_e\,.
\vspace*{1.5mm}
\label{eq:ampl1}
\ee 
Here $G_\beta\equiv G_F V_{ud}$, $G_F$ and $V_{ud}$ being the Fermi constant 
and the $ud$ element of the CKM matrix, $U_{ej}$ are the elements of the 
leptonic mixing matrix, $\varepsilon_\mu$ is the polarization 4-vector of the 
parent nucleus, $M_{\rm GT}$ is the nuclear matrix element, $F(Z, E_e)$ is 
the Fermi function which takes into account the interaction 
of the produced electron with the Coulomb field of the daughter nucleus, and 
$E_j$ and $E_e$ are the neutrino and electron energies. Energy conservation 
yields
\be
E_e=E_j+E_0\,,
\label{eq:encons}
\ee
where $E_0$ is the total energy release in the corresponding $\beta^-$-decay 
process, which is related to the usually quoted $Q_\beta$-value of the process 
in the limit of vanishing neutrino mass by $E_0=Q_\beta+m_e$. 
The leptonic tensor $X^{\mu\nu}$ is 
\be
X^{\mu\nu}=
[\bar{u}_{e}(k)\gamma^\mu(1-\gamma_5)u_{j}(q)]
[\bar{u}_{j}(q)\gamma^\nu(1-\gamma_5)u_{e}(q)]\,,
\label{eq:X1}
\ee
where $u_e(k)$ is the electron spinor and $u_j(q)$ is that of the neutrino mass 
eigenstate with mass $m_j$. For electron and neutrino in definite spin states  
we have 
\bea
&&u_e(k)\bar{u}_e(k)=\frac{1}{2}(\slashed{k}+m_e)(1+\gamma_5\slashed{S}_e)\,,
\qquad
\label{eq:eproj}
\\
&&u_{j}(q)\bar{u}_{j}(q)=\frac{1}{2}(\slashed{q}+m_j)(1+\gamma_5
\slashed{S}_{j})\,,
\vspace*{-1.5mm}
\label{eq:nuproj}
\vspace*{-1.5mm}
\eea 
where $m_e$ and $m_j$ are the electron and neutrino masses, and 
$S_e^\mu$ and $S_j^\mu$ are their spin 4-vectors:
\be
S_e^\mu
=\bigg(\frac{\vec{k}\!\cdot\!\vec{s}_e}{m_e},\;\vec{s}_e+\frac{
(\vec{k}\!\cdot\!\vec{s}_e)\vec{k}}{m_e(E_e+m_e)}\bigg)\,,
\label{eq:Se}
\ee
\be
S_j^\mu=\bigg(\frac{\vec{q}\!\cdot\!\vec{s}_j}{m_j},\;\vec{s}_j+\frac{
(\vec{q}\!\cdot\!\vec{s}_j)\vec{q}}{m_j(E_j+m_j)}\bigg)\,.
\label{eq:Sj}
\ee
Here $\vec{s}_e$ and $\vec{s}_j$ are, respectively, the unit vectors in the 
direction of the electron and neutrino spin in their rest frames.  
Introducing the 4-vectors
\be
A^\mu\equiv k^\mu-m_e S_e^\mu\,,\qquad\quad B^\mu\equiv q^\mu-m_j S_j^\mu\,,  
\label{eq:AB1}
\ee
 we can write the squared 
amplitude $\varepsilon_{\mu}\varepsilon_{\nu}^* X^{\mu\nu}$ in a very 
compact form:%
\footnote{
\label{fn:note1}
Note that, since the electron and neutrino spin 4-vectors satisfy the usual 
relations $k \cdot S_e=q \cdot S_j=0$ and $S_e^2=S_j^2=-1$, the 4-vectors 
$A^\mu$ and $B^\mu$ are lightlike: $A^2=B^2=0$. 
When the vector $\vec{s}_j$ is aligned or antialigned with 
$\vec{s}_N$, for neutrinos that are in helicity eigenstates  
the vector $\vec{B}$ is also aligned or antialigned with $\vec{s}_N$. 
In this case from $B^2=0$ it follows that $\vec{B}\!\cdot\!\vec{s}_N=
\pm B^0$. For $\vec{s}_j\parallel \vec{s}_N$ we find 
$\vec{B}\!\cdot\!\vec{s}_N=-B^0$, and the transition amplitude vanishes. One 
can similarly show that it also vanishes when the emitted electron is in a 
helicity eigenstate and its spin is antialigned with $\vec{s}_N$ 
(see also section~\ref{sec:disc} below).}
\be
\varepsilon_{\mu}\varepsilon_{\nu}^* X^{\mu\nu}=
2\big(A^0-\vec{A}\!\cdot\!\vec{s}_N\big)
\big(B^0+\vec{B}\!\cdot\!\vec{s}_N\big)\,.
\label{eq:product2}
\vspace*{1.5mm}
\ee
Here $\vec{s}_N$ is the unit vector in the direction of the spin of the 
initial-state nucleus. Defining 
\be
K_e\equiv 
1-\frac{E_e}{E_e+m_e}\vec{v}_e\!\cdot\!\vec{s}_e
\,,\qquad\quad 
K_j \equiv 
1-\frac{E_j}{E_j+m_j}\vec{v}_j\!\cdot\!\vec{s}_j
\,, 
\label{eq:KK}
\ee
we can rewrite $A^\mu$ and $B^\mu$ as 
\bea
A^\mu= E_e\Big(1-\vec{v}_e\!\cdot\!\vec{s}_e
,\;K_e\vec{v}_e-
\frac{m_e}{E_e}\vec{s}_e\Big)\,,
\label{eq:AB3a}
\\
B^\mu= E_j\Big(1-\vec{v}_j\!\cdot\!\vec{s}_j,\;K_j\vec{v}_j-
\frac{m_j}{E_j}\vec{s}_j\Big)\,. 
\label{eq:AB3b}
\eea
Substituting these expressions 
into eq.~(\ref{eq:product2}), we find 
\bea
\varepsilon_{\mu}\varepsilon_{\nu}^* X^{\mu\nu}=2E_e E_j
\big[(1-\vec{v}_e\!\cdot\!\vec{s}_e)-K_e(\vec{v}_e\!\cdot\!\vec{s}_N)
+\frac{m_e}{E_e}
(\vec{s}_e\!\cdot\!\vec{s}_N)
\big]
\nonumber \\
\times \big[(1-\vec{v}_j\!\cdot\!\vec{s}_j)+K_j(\vec{v}_j\!\cdot\!\vec{s}_N)
-\frac{m_j}{E_j}(\vec{s}_j\!\cdot\!\vec{s}_N)\big].
\label{eq:product3}
\vspace*{-1.5mm}
\eea

The above expressions give the differential cross section of neutrino capture 
on polarized nuclei (in the particular case of allowed $1^\pi\to 
0^\pi$ Gamow-Teller transitions) without summation over the spin states  
of any of the involved particles or integration over the directions of their 
momenta. To our knowledge, no such expression has been previously derived. 
When the summations over the spin states of the neutrino and the electron, 
or of the neutrino and the parent nucleus, or the summation over the 
neutrino spin states and integration over the direction of its momentum are 
performed, our results coincide with the corresponding results found by 
Jackson et al. \cite{jtw1,jtw2}.

The amplitude of the process under consideration depends on five vectors -- 
$\vec{v}_e$, $\vec{v}_j$, $\vec{s}_e$, $\vec{s}_j$ and $\vec{s}_N$, 
from which one can form ten scalar dot-products:
\begin{align}
&
\vec{v}_e\cdot \vec{s}_e\,,\quad \vec{v}_e\cdot \vec{s}_N\,,\quad
\vec{s}_e\cdot \vec{s}_N\,,\quad 
\vec{v}_j\cdot \vec{s}_j\,,
\nonumber \\
&
\vec{v}_e\cdot \vec{v}_j\,,\quad
\vec{v}_e\cdot \vec{s}_j\,,\quad
\vec{v}_j\cdot \vec{s}_e\,,\quad
\vec{s}_e\cdot \vec{s}_j\,,\quad
\vec{v}_j\cdot \vec{s}_N\,,\quad
\vec{s}_j\cdot \vec{s}_N\,.
\label{eq:scal2}
\end{align}
Eq.~(\ref{eq:product3}) actually contains only on six of them: 
terms containing the quantities $\vec{v}_e\cdot \vec{v}_j$, $\vec{v}_e\cdot 
\vec{s}_j$, $\vec{v}_j\cdot \vec{s}_e$ and $\vec{s}_e\cdot \vec{s}_j$ are 
absent from the squared amplitude of the process. They will, however, arise 
if one averages over the directions of nuclear polarization $\vec{s}_N$, 
i.e.\ considers neutrino capture on unpolarized nuclei (see 
eqs.~(\ref{eq:prodNew})-(\ref{eq:prodNew2d}) in 
section~\ref{sec:betanu} below).  

Out of the ten scalar dot-products in eq.~(\ref{eq:scal2}), those in the first 
line are not useful for discriminating between neutrino capture and 
neutrino emission in $\beta$ decay. Indeed, the first three of them do not 
depend on the neutrino variables, whereas the last one, $\vec{v}_j\cdot 
\vec{s}_j$, is not helpful because experiments on neutrino helicity 
measurements cannot distinguish between absorption of a 
left-helical neutrino and production of a right-helical (anti)neutrino. 
The term $\vec{v}_j\cdot\vec{s}_j$, however, affects the total neutrino 
capture rate and may also play an important role in studying Dirac vs.\ 
Majorana neutrino nature in captures of relic neutrinos 
(see \cite{long,roulet} and section~\ref{sec:movframe} below).

Out of the six scalars in the second line of (\ref{eq:scal2}), only the last 
two enter into the squared amplitude in eq.~(\ref{eq:product3}). 
They result in angular correlations between the nuclear 
polarization and the directions of neutrino spin and velocity. 
Summing eq.~(\ref{eq:product3}) over the electron spin states and averaging 
over the directions of the produced electrons, we find 
\be
\int\frac{d\Omega_e}{4\pi}
\sum_{s_e}\varepsilon_{\mu}\varepsilon_{\nu}^* X^{\mu\nu}=4E_e E_j
\Big\{(1-\vec{v}_j\!\cdot\!\vec{s}_j)
+\big(K_j\vec{v}_j-\frac{m_j}{E_j}\vec{s}_j\big)\!\cdot\!\vec{s}_N
\Big\}.
\label{eq:product5}
\ee 
If the direction of nuclear polarization is fixed in the Earth frame, this 
squared amplitude will vary with time because of variations of the 
directions of the spin and velocity of the incoming neutrinos. 
Therefore, in experiments with polarized nuclear targets one can study 
time variations of the relic neutrino signal by merely measuring the total 
rate of electron production. At the same time, as we shall see in 
section~\ref{sec:betanu}, measuring the angular and/or spin distributions 
of the produced electrons will be useful in the case of experiments with 
unpolarized nuclei. 

In order to derive the expressions for the total capture rate of the 
C$\nu$B neutrinos as well as for various angular correlations of interest, 
one has to multiply eq.~(\ref{eq:ampl1}) by the neutrino velocity 
distribution function of the $j$th neutrino mass-eigenstate 
$f(\vec{v}_j)$, integrate or sum over the relevant finite-state kinematic 
variables and sum the result over $j$.

\subsection{\label{sec:betanu}Neutrino capture on unpolarized nuclei}

Polarizing sufficiently large targets and maintaining (or renewing) their 
polarization during extended intervals of time may pose substantial 
experimental difficulties. Can angular correlations help discriminate 
between relic neutrino capture and $\beta$ decay of the target nuclei in 
experiments with unpolarized targets? One possibility is to 
make use of spin polarization of the daughter nuclei with $J\ne 0$. 
This polarization can be studied, for instance, 
by measuring circular polarization of the de-excitation $\gamma$-quanta 
in $\beta$-transitions into excited states of the daughter nuclei, as in 
the famous experiment of Goldhaber, Grodzins and Sunyar \cite{goldh}. For 
the particular case of the allowed Gamow-Teller $0^\pi \to 1^\pi$ 
transitions, the corresponding squared amplitude can be obtained from the 
expressions in eqs.~(\ref{eq:product2}) or (\ref{eq:product3}) by the 
substitution $\vec{s}_N\to -\vec{s}_N$ (assuming that the velocity of the 
recoil nucleus can be neglected).  

Another, probably more practical, possibility is to study angular 
distributions of the produced electrons or their spin states. For neutrino 
capture on unpolarized nuclei in allowed Gamow-Teller $1^\pi \to 0^\pi$ 
transitions the squared amplitude can be found by averaging 
$\varepsilon_{\mu}\varepsilon_{\nu}^*X^{\mu\nu}$ over the polarizations 
$\lambda$ of the parent nucleus. Direct calculation gives  
\be
\frac{1}{3}\sum_\lambda
\varepsilon_{\mu}(\lambda)\varepsilon_{\nu}^*(\lambda)
X^{\mu\nu}
=2\,\big[A^0 B^0 -\frac{1}{3}\vec{A}\!\cdot\!\vec{B}\,\big].
\label{eq:prodNew}
\ee
Note that the same result can be obtained by averaging eq.~(\ref{eq:product2}) 
over the directions of $\vec{s}_N$ (i.e.\ by taking 
$\int (d\Omega_{\vec{s}_N}/4\pi)$ of both its parts). Let us first consider 
the case when the spin state of the produced electron is not measured. 
Summing eq.~(\ref{eq:prodNew}) over $s_e$, we find 
\be
\frac{1}{3}\sum_{\lambda, s_e}
\varepsilon_{\mu}(\lambda)\varepsilon_{\nu}^*(\lambda)
X^{\mu\nu}=4E_e\big(B^0 -\frac{1}{3}\vec{B}\!\cdot\!\vec{v}_e\big)\,,
\label{eq:prodNew2a}
\ee
or, in a more detailed form, 
\begin{align}
\frac{1}{3}\sum_{\lambda, s_e}
\varepsilon_{\mu}(\lambda)\varepsilon_{\nu}^*(\lambda)
X^{\mu\nu}
=4E_eE_j\,\Big[1-\vec{v}_j\!\cdot\!\vec{s}_j-\frac{1}{3}\vec{v}_e
\!\cdot\!\vec{v}_j+\frac{1}{3}
\frac{E_j}{E_j+m_j}(\vec{v}_j\!\cdot\!\vec{s}_j)
(\vec{v}_e\!\cdot\!\vec{v}_j)+\frac{1}{3}
\frac{m_j}{E_j}\vec{v}_e\!\cdot\!\vec{s}_j
\Big].
\label{eq:prodNew2c}
\end{align}
Alternatively, one can consider the situation when the direction of the spin 
of the produced electrons is observed, while the directions of their momenta 
are not. Then the relevant squared amplitude is 
\[
\int
\frac{d\Omega_e}{4\pi}\,
\frac{1}{3}\sum_{\lambda}
\varepsilon_{\mu}(\lambda)\varepsilon_{\nu}^*(\lambda)X^{\mu\nu}
=2E_e\Big\{B^0+\frac{1}{9}\Big(1+2\frac{m_e}{E_e}\Big)
(\vec{s}_e\!\cdot\!\vec{B})\Big\} 
\hspace*{4.5cm}
\]
\be
\qquad=2E_e E_j\Big\{ 1-\vec{v}_j\!\cdot\!\vec{s}_j 
+\frac{1}{9}\Big(1+2\frac{m_e}{E_e}\Big)
\Big[\vec{v}_j\!\cdot\!\vec{s}_e-\frac{E_j}{E_j+m_j}
(\vec{v}_j\!\cdot\!\vec{s}_j)(\vec{v}_j\!\cdot\!\vec{s}_e)-\frac{m_j}{E_j}
\vec{s}_e\!\cdot\!\vec{s}_j\Big]
\Big\}. 
\label{eq:prodNew2d}
\ee

Eq.~(\ref{eq:prodNew2c}) describes anisotropies of electron emission with 
respect to the directions of the velocity and spin of the incoming relic 
neutrinos $\vec{v}_j$ and $\vec{s}_j$, which change with time
in the lab frame. The electron direction anisotropy with respect to a 
fixed direction in this frame should therefore exhibit time variations. 
The same applies to the electron spin anisotropy 
described by eq.~(\ref{eq:prodNew2d}). 
This in principle could be used to find out the direction 
of the peculiar motion of the Sun with respect to the C$\nu$B rest 
frame. 

In the case when neither the spin state nor the direction of the produced 
electron is observed, the relevant squared amplitude is
\be
\int\frac{d\Omega_e}{4\pi}\,\frac{1}{3}\sum_{\lambda,s_e}
\varepsilon_{\mu}(\lambda)\varepsilon_{\nu}^*(\lambda)X^{\mu\nu}
=4E_e E_j\,\big(1-\vec{v}_j\!\cdot\!\vec{s}_j\big)\,,
\vspace*{-2.0mm}
\label{eq:prodNew2e}
\ee
which is the usual inclusive squared amplitude for neutrino 
capture on unpolarized nuclei.

\subsection{\label{sec:trit}Comparison with 
$\nu$ capture on polarized tritium ($1/2^+\to1/2^+$)
}

It is instructive to compare our results obtained for pure Gamow-Teller 
$1^\pi\to 0^\pi$ transitions with those found by Lisanti 
et al.\ for neutrino capture on polarized tritium, which is the mixed 
Fermi--Gamow-Teller $1/2^+\to1/2^+$ transition \cite{lisanti1}. 
The authors calculated the squared amplitude in the case when the spin 
states of the produced electron and daughter $^3$He are not measured and 
found the angular correlation  (in our notation)  
\be
1-\vec{v}_j\!\cdot\vec{s}_j + A(1-\vec{v}_j\!\cdot\vec{s}_j)
\vec{v}_e\!\cdot\vec{s}_N +B K_j \vec{v}_j\!\cdot\vec{s}_N 
-B\frac{m_j}{E_j}\vec{s}_j\!\cdot\vec{s}_N
+a K_j \vec{v}_e\!\cdot\vec{v}_j 
-a\frac{m_j}{E_j}\vec{v}_e\!\cdot\vec{s}_j\,.
\label{eq:compare1}
\ee
Here $A$, $B$ and $a$ are the standard angular correlation coefficients 
\cite{jtw1,jtw2}, which can be expressed through the ratios of the Fermi and 
Gamow-Teller nuclear matrix elements $M_{\rm F}$ and $M_{\rm GT}$. For the 
considered case of neutrino capture 
on polarized tritium their numerical values are \cite{lisanti1} 
\be
A\simeq -0.095\,,\qquad\quad B\simeq 0.99\,,\qquad\quad
a\simeq -0.087\,.\qquad\quad
\label{eq:num1}
\ee
To compare eq.~(\ref{eq:compare1}) with our results, one has to sum our 
squared amplitude 
(\ref{eq:product3}) over $s_e$. The obtained angular correlation is 
\be
1-\vec{v}_j\!\cdot\vec{s}_j-
(1-\vec{v}_j\!\cdot\vec{s}_j)\vec{v}_e\!\cdot\vec{s}_N 
+K_j \vec{v}_j\!\cdot\vec{s}_N
-\frac{m_j}{E_j}\vec{s}_j\!\cdot\vec{s}_N
-K_j(\vec{v}_j\!\cdot\vec{s}_N)(\vec{v}_e\!\cdot\vec{s}_N)
+\frac{m_j}{E_j}(\vec{v}_e\!\cdot\vec{s}_N)(\vec{s}_j\!\cdot\vec{s}_N)\,.
\label{eq:compare2}
\ee
One can see that the first five terms in eq.~(\ref{eq:compare1})  
have the same structure as the first five terms in eq.~(\ref{eq:compare2}); 
the two sets coincide with each other if one chooses $A=-1$ and $B=1$ 
in~(\ref{eq:compare1}). At the same time, the last 
two terms in (\ref{eq:compare1}), which have the coefficient $a$ as 
a factor, are different from the last two terms in (\ref{eq:compare2}). 
The latter are bilinear in the polarization vector of the parent nucleus 
$\vec{s}_N$, while eq.~(\ref{eq:compare1}) contains only terms of zero and 
first power in $\vec{s}_N$. The origin of this difference in the structures 
of eqs.~(\ref{eq:compare1}) 
and (\ref{eq:compare2}) lies in the special nature of the $1/2^+\to1/2^+$ 
transition, for which the tensor alignment 
$\propto [J(J+1)-3\langle(\vec{J}\!\cdot\!\vec{s}_N)^2\rangle]$ 
vanishes \cite{jtw1}. 
It is interesting to note that upon averaging over the directions of 
$\vec{s}_N$ the last two terms in eq.~(\ref{eq:compare2}) would reproduce 
those in eq.~(\ref{eq:compare1}) with $a=-1/3$.

Although effects of angular correlations 
are in general more prominent 
in the case of pure Gamow-Teller transitions than in the case of mixed 
Fermi--Gamow-Teller ones, 
the correlation between the velocity (and/or spin) of the incoming neutrino 
and nuclear polarization in neutrino capture on polarized tritium is quite 
substantial. This follows from the fact that the coefficient $B$ 
governing these correlations is close to unity (see eq.~(\ref{eq:num1})). 
At the same time, the $\beta-\nu$ correlation, which exists even for 
processes on unpolarized nuclei, is suppressed in this transition due to 
the numerical smallness of the coefficient $a$. By contrast, this 
coefficient is significant in the case of pure transitions: $a=-1/3$ for 
pure Gamow-Teller allowed transitions and $a=1$ for pure Fermi allowed or 
superallowed transitions $0^\pi\to 0^\pi$.

\section{\label{sec:movframe} 
Implications for relic neutrino capture: lab frame}

Standard cosmology predicts the existence of a background of  nearly 
uniform and isotropic cosmic neutrinos with the average density of 
$n_{\nu 0}\simeq 56$ cm$^{-3}$ per mass eigenstate and per spin degree of 
freedom. Relic neutrinos are expected to be in helicity eigenstates in 
the C$\nu$B rest frame, and therefore  not only their velocities, but 
also their spin directions should be distributed isotropically in that 
frame. As at least two relic neutrino species should now be 
non-relativistic, gravitational clustering may modify the local density 
of C$\nu$B at the Earth's location (see, e.g., \cite{salas} and 
references therein). Gravitational focusing by the Sun can also modify 
local neutrino density and velocity distribution, leading to annual 
modulations of the relic neutrino signal at the Earth \cite{safdi}. 
These effects are, however, expected to be relatively small in the case of 
hierarchical neutrino mass spectrum and will not be discussed here. 

The Sun is expected to have a non-zero peculiar velocity with respect to the 
C$\nu$B rest frame, therefore there should exist a ``wind'' of relic neutrinos 
at the Earth, i.e.\ they should have a preferred direction of arrival in the 
Earth's rest frame. 
Although the speed and the direction of the peculiar motion of the Sun are 
model-dependent and thus not precisely known, it is expected that the 
peculiar velocity $u$ is rather small, ${\cal O}(10^{-3})$. 
The daily rotation of the Earth about its axis means that direction of the 
neutrino ``wind'' changes during the day in the lab frame, which should lead 
to modulations of various angular correlations in relic neutrino capture  
on both polarized and unpolarized nuclei. Our goal is to study if these 
modulations can be used to distinguish neutrino capture 
from the usual $\beta$-decay of target nuclei. 
Our discussion of the effects of anisotropy of relic neutrinos 
will differ from that in ref.~\cite{lisanti1} in two important respects: 

\begin{itemize}
\item While the authors of \cite{lisanti1} concentrated  on 
non-relativistic relic neutrinos, we consider the general case, since  
in the case of the normal neutrino mass ordering currently preferred by 
the data the lightest neutrino mass eigenstate $\nu_1$ (which has the largest 
contribution of the electron flavour 
$|U_{e1}|^2\simeq 0.66$) may still be relativistic.

\item
We consider angular correlations in neutrino captures on both 
polarized and unpolarized nuclear targets. In particular, the $\beta-\nu$ 
angular correlation can be quite sizeable for unpolarized pure Gamow-Teller 
or pure Fermi transitions though it is suppressed for the mixed 
Fermi--Gamow-Teller one studied in \cite{lisanti1}. 

\end{itemize}

We want to find the neutrino differential capture rate in an Earth-bound lab 
frame which moves with respect to the C$\nu$B rest frame with the velocity 
$-\vec{u}$ with $u\equiv|\vec{u}|\ll 1$. To this end, we consider the effects 
of the Lorentz boost on the neutrino variables to first order in the boost 
velocity $u$. 
Let us denote neutrino variables in the lab frame with primed quantities, 
whereas the unprimed variables will refer to the C$\nu$B rest frame. 
For the energy, momentum and velocity of the $j$th neutrino mass eigenstate 
in the lab frame we then have 
\be
E_j'=E_j+\vec{u}\!\cdot\!\vec{q}\,,\qquad
\vec{q}\,'=\vec{q}+\vec{u}E_j\,,\qquad
\vec{v}\,'\!\!_j=\vec{v}_j(1-\vec{u}\!\cdot\!\vec{v}_j)+\vec{u}\,.
\label{eq:Ep}
\ee
We will be mostly interested in the regime 
$u\ll v_j$, which in the standard 3-flavour neutrino picture corresponds 
to hierarchical neutrino mass spectrum with the largest neutrino mass 
of order $0.05$ eV. In this limit for the unit vector 
in the direction of the neutrino velocity in the Earth frame we find 
\be
\frac{\vec{v}\,'\!\!_j}{v_j'}=
\frac{1}{v_j}\Big\{\vec{v}_j+\big[\vec{u}-
\frac{(\vec{u}\!\cdot\!\vec{v}_j)}{v_j^2}\vec{v}_j\big]\Big\}\equiv 
\frac{1}{v_j}\big(\vec{v}_j+\vec{u}_{\perp j}
\big)\,,
\vspace*{1.5mm}
\label{eq:vv}
\ee
where $\vec{u}_{\perp j}$ is the component of the boost velocity $\vec{u}$ 
orthogonal to $\vec{v}_j$.

The neutrino spin requires a special consideration. The spin 4-vector 
$S_j^\mu$ in eq.~(\ref{eq:Sj}) can be obtained from the rest-frame vector 
$S_j^{(0)\mu}=(0,\vec{s}_j)$ by a Lorentz boost. The corresponding 4-vector 
in the Earth frame, $S_j'^{\mu}$, is then obtained by another boost, with 
velocity $\vec{u}$. The result will have the form similar to  
that in eq.~(\ref{eq:Sj}), with all the quantities in it replaced by the primed 
ones. Note that in general $\vec{s}\,'\!\!_j\neq \vec{s}_j$; this is related 
to the fact that a sequence of two boosts with non-collinear velocities is 
equivalent to a boost and a rotation (called the Wigner rotation) rather than 
to a single boost. The general expression for $\vec{s}\,'\!\!_j$ is rather 
involved, but to first order in $u$ one finds a 
simple result%
\footnote{It can be readily obtained by applying the Lorentz 
boost to linear order in the boost velocity $\vec{u}$ to the 4-vector 
$S_j^\mu$ and requiring that in the primed variables it have the same form 
as eq.~(\ref{eq:Sj}).}  
\be
\vec{s}\,'\!\!_j=\vec{s}_j+\frac{(\vec{q}\!\cdot\!\vec{s}_j)\vec{u}-
(\vec{u}\!\cdot\!\vec{s}_j)\vec{q}}{E_j+m_j}\,. 
\label{eq:s}
\ee
Assume now that in the C$\nu$B rest frame neutrinos are in the exact helicity 
eigenstates, i.e.\ 
\be
\vec{s}_j=\lambda_j \frac{\vec{v}_j}{v_j}\,, 
\label{eq:helic1}
\ee
where $\lambda_j=\pm 1$, with the upper (lower) sign corresponding to 
right-helical (left-helical) neutrinos. Then 
\be
\vec{s}\,'\!\!_j=\lambda_j\Big\{
\frac{\vec{v}_j}{v_j}+v_j\frac{E_j}{E_j+m_j}\vec{u}_{\perp j}\Big\}.
\label{eq:s2} 
\ee
It is interesting to note that, even though helicity is not a 
Lorentz-invariant quantity, in the regime $u\ll v_j$ of main interest to us 
and to first order in the boost velocity $u$ the neutrino helicity in the lab 
frame is the same as in the C$\nu$B rest frame. This follows from the fact 
that for small boosts the correction to $\vec{s}_j$ (i.e.\ the second term in 
the curly brackets in eq.~(\ref{eq:s2})) is of order $u$ and orthogonal to 
$\vec{v}_j$, and in the limit $u\ll v_j$ the correction to the unit vector of 
the neutrino velocity in eq.~(\ref{eq:vv}) is orthogonal to 
$\vec{s}_j$, which is collinear with $\vec{v}_j$. 
Therefore, the correction to $\vec{s}_j\cdot\vec{v}_j/v_j$ is quadratic in 
$u$ in this limit and can be neglected.%
\footnote{
Note that this would not be in general correct for $u\gtrsim v_j$ (which 
corresponds to neutrino mass $m_j\gtrsim 0.5$ eV), even if $u$ is small. 
In particular, for $v_j\ll u\ll 1$ we would have  
$\vec{v}\,'\!\!_j\simeq \vec{u}$, 
$\vec{s}\,'\!\!_j\simeq\vec{s}_j$ and 
$\vec{s}\,'\!\!_j\cdot\vec{v}_j'/v_j' \simeq 
\vec{s}_j\cdot\vec{u}/u$, which is different from $\lambda_j$ for all boost 
velocities $\vec{u}$ that are not parallel or antiparallel to $\vec{v}_j$.  
}

Next, we will consider the consequences of the fact that C$\nu$B is isotropic 
in its rest frame for the average characteristics of relic neutrinos in the 
lab frame. In the C$\nu$B rest frame we have    
$\langle\vec{v}_j\rangle=0$, $\langle\vec{s}_j\rangle=0$ (hereafter the 
angular brackets will denote averaging over the angular distributions of the 
C$\nu$B neutrinos but not over the absolute values of the neutrino velocities 
$v_j$). 
Therefore, for the quantities in the lab frame we get 
\be
\langle E_j'\rangle=E_j\,,\qquad \langle\vec{q}\,'\rangle=
\vec{u} E_j\,,\qquad \langle \vec{v}\,'\!\!_j\rangle=\Big(1-\frac{v_j^2}{3}\Big)
\vec{u}\,.
\label{eq:Epaver}
\ee
Further discussion of the effects of averaging over the directions of C$\nu$B 
neutrinos on neutrino observables in the lab frame can be found in Appendix A.

\subsection{\label{sec:pol}Polarized targets}

Let us consider now C$\nu$B detection by capture on polarized targets in the 
lab frame. For the squared amplitude of the allowed $1^\pi\to 0^\pi$ 
transition one finds (see Appendix A)
\be
\Big\langle
\frac{\varepsilon_{\mu}\varepsilon_{\nu}^* X^{\mu\nu}}{4E_e E_j'}\Big\rangle=
\frac{1}{2E_e}\big(A^0-\vec{A}\!\cdot\!\vec{s}_N\big)
\Big\{1-\lambda_j v_j+\Big(1-\frac{2}{3}\lambda_j v_j-\frac{v_j^2}{3}\Big)
\vec{u}\!\cdot\!\vec{s}_N
\Big\}.
\label{eq:factor1}
\ee
The part of this expression relevant 
to our discussion is the last factor on the right hand side. 
As the direction of $\vec{u}$ in the lab frame changes during the day 
because of the rotation of the Earth, this factor 
gives rise to periodic variations of the signal,  
provided that the nuclear polarization vector $\vec{s}_N$ is not 
oriented along the Earth's rotation axis. The amplitude of the time 
variations is maximal when $\vec{s}_N$ is orthogonal to this axis.%
\footnote{In the geocentric spherical coordinates 
we have $\vec{u}\!\cdot\!\vec{s}_N=u[\cos\theta_u
\cos\theta_N+\sin\theta_u\sin\theta_N\cos(\phi_u(t)-\phi_N)]$ with 
$\phi_u(t)=(2\pi/T_0)t+\phi_0$, $T_0$ being the sidereal day. 
As the declination $\theta_u$ of the vector $\vec{u}$ is fixed, the 
time dependence of $\vec{u}\!\cdot\!\vec{s}_N$ is maximized when 
$\sin\theta_N=1$, i.e.\ when $\vec{s}_N$ is orthogonal to the Earth's 
rotation axis.} 
Let us now examine the effects of these time-dependent angular correlations. 

Let ${\cal F}_j(\lambda_j)$ denote the expression in the curly brackets in 
eq.~(\ref{eq:factor1}). 
For left-helical neutrinos 
we have 
\be
{\cal F}_j(\lambda_j=-1)=
1+v_j+
\Big(1+\frac{2}{3}v_j-\frac{v_j^2}{3}
\Big)\vec{u}\!\cdot\!\vec{s}_N\,.
\label{eq:factor1a}
\ee
For right-helical neutrino states we have to distinguish between 
Dirac and Majorana neutrinos. In the Dirac case, right-helical states 
are antineutrinos which cannot be detected in inverse $\beta^-$-decay 
processes, that is, one has to set ${\cal F}_j(\lambda_j=1)=0$ in that case. 
For Majorana neutrinos we have 
\be
{\cal F}_j(\lambda_j=1)=
1-v_j+
\Big(1-\frac{2}{3}v_j-\frac{v_j^2}{3}
\Big)\vec{u}\!\cdot\!\vec{s}_N\,.
\label{eq:factor1c}
\ee
In the limit $v_j\to 1$ this expression vanishes, 
in accord with the fact that 
ultra-relativistic Dirac and Majorana neutrinos are essentially 
indistinguishable.

Assuming that C$\nu$B contains equal numbers of left-helical and right-helical 
neutrinos and summing over the helicities, we find 
\be
{\cal F}_j\equiv 
\sum_{\lambda_j=\pm 1}{\cal F}_j(\lambda_j)=
\left\{
\begin{array}{ll}
~\,1+v_j+
\Big(1+\frac{2}{3}v_j-\frac{v_j^2}{3}
\Big)\vec{u}\!\cdot\!\vec{s}_N\,,& 
\mbox{Dirac neutrinos}\,,
\vspace*{2.5mm}
\\
2\Big[
1+
\Big(1-\frac{v_j^2}{3}
\Big)\vec{u}\!\cdot\!\vec{s}_N\Big]\,,& \mbox{Majorana neutrinos}\,.
\end{array}
\right.
\label{eq:factor2}
\ee
In the limiting case of non-relativistic neutrinos this gives 
\be
{\cal F}_j
\simeq\left\{
\begin{array}{ll}
~\,1+v_j+\big(1+\frac{2}{3}v_j\big)\vec{u}\!\cdot\!\vec{s}_N \,,& 
\mbox{Dirac neutrinos}\,,
\vspace*{2.5mm}
\\
2\Big(
1+\vec{u}\!\cdot\!\vec{s}_N\Big)\,,& \mbox{Majorana neutrinos}\,.
\end{array}
\right.
\label{eq:factor2a}
\ee
For highly relativistic neutrinos we obtain
\be
{\cal F}_j^{Dir}\simeq{\cal F}_j^{Maj}\simeq
2\Big(1+\frac{2}{3}\vec{u}\!\cdot\!\vec{s}_N\Big).
\label{eq:factor2b}
\ee

Let us consider more closely the regime of non-relativistic neutrinos, in which 
the results for Dirac and Majorana neutrinos differ. Dropping in  
(\ref{eq:factor2a}) the term proportional to $\vec{u}\!\cdot\!\vec{s}_N$, 
we recover the result of \cite{long} obtained for neutrino absorption by 
unpolarized nuclei: the detection cross section for Majorana 
neutrinos is about twice as large as that for Dirac neutrinos, which means that 
detection of relic neutrinos could in principle shed light on neutrino nature. 
This is, however, complicated by the fact that the local C$\nu$B density at the 
Earth may differ from $n_{\nu 0}$ due to gravitational clustering effects and 
so is not precisely known. Because the detection rate of relic neutrinos 
is proportional to their local density, by measuring only the absolute 
C$\nu$B detection rate one may not be able to determine neutrino nature 
unambiguously. 

As can be seen from eq.~(\ref{eq:factor2a}), this degeneracy between unknown 
local relic neutrino density and Dirac/Majorana neutrino nature can in 
principle be lifted by measuring time-dependent angular correlations in relic 
neutrino capture. This follows from the difference of the dependences of 
${\cal F}_j$ on $\vec{u}\!\cdot\!\vec{s}_N$ in the Dirac and Majorana cases, 
which cannot be absorbed in the overall normalization of ${\cal F}_j$. 
In practice, however, making use of this difference will be very difficult 
because it contains an extra factor $v_j$ compared to the main   
$\vec{u}\!\cdot\!\vec{s}_N$ term. 

\subsection{\label{sec:unpol}Unpolarized targets}

Let us now turn to angular correlations in the case of relic neutrino 
capture on unpolarized targets. By averaging eq.~(\ref{eq:factor1}) 
over the polarizations of the parent nuclei we find  
\be
\Big\langle
\frac{1}{3}\sum_\lambda
\frac{\varepsilon_{\mu}(\lambda)\varepsilon_{\nu}^*(\lambda) X^{\mu\nu}}
{4E_e E_j'}\Big\rangle=
\frac{1}{2E_e}\Big\{A^0(1-\lambda_j v_j)-\frac{1}{3}
\Big(1-\frac{2}{3}\lambda_j v_j-\frac{v_j^2}{3}\Big)
\vec{u}\!\cdot\!\vec{A}\Big\}.
\label{eq:factor3}
\ee
Consider this expression in two special cases. If the spin state of the 
produced electron is not measured, summing over $s_e$ we obtain
\be
\Big\langle
\frac{1}{3}\sum_{\lambda, s_e}
\frac{\varepsilon_{\mu}(\lambda)\varepsilon_{\nu}^*(\lambda) X^{\mu\nu}}
{4E_e E_j'}\Big\rangle=
1-\lambda_j v_j-\frac{1}{3}
\Big(1-\frac{2}{3}\lambda_j v_j-\frac{v_j^2}{3}\Big)
\vec{u}\!\cdot\!\vec{v}_e\,.
\label{eq:factor3a}
\ee
If the electron direction is not observed but its spin state is, we find 
\be
\int\frac{d\Omega_e}{4\pi}
\Big\langle
\frac{1}{3}\sum_{\lambda}
\frac{\varepsilon_{\mu}(\lambda)\varepsilon_{\nu}^*(\lambda) X^{\mu\nu}}
{4E_e E_j'}\Big\rangle=
\frac{1}{2}
\Big\{1-\lambda_j v_j+\frac{1}{9}
\Big(1+2\frac{m_e}{E_e}\Big)
\Big(1-\frac{2}{3}\lambda_j v_j-\frac{v_j^2}{3}\Big)
\vec{u}\!\cdot\!\vec{s}_e \Big\}.
\label{eq:factor3b}
\ee
Through the terms proportional to $\vec{u}\cdot\vec{v}_e$ and 
$\vec{u}\cdot\vec{s}_e$, these expressions exhibit angular correlations 
between the directions of the electron's velocity 
or spin and the preferred direction of relic 
neutrino arrival $\vec{u}$ which changes in the lab frame during the day. 
Therefore, the probability of electron emission with its momentum (or spin) 
pointing in a certain direction 
in the lab frame will in general exhibit periodic time variations. 
The variations are absent for the directions collinear with the Earth's 
rotation axis and are maximized for the orthogonal directions (see Appendix B). 

Note that the right hand side of eq.~(\ref{eq:factor3a}) can be formally 
obtained from ${\cal F}_j(\lambda_j)$ 
by the replacement  
$\vec{u}\cdot\vec{s}_N\to (-1/3)\vec{u}\cdot\vec{v}_e$,  and 
similarly that of eq.~(\ref{eq:factor3b}) is obtained from 
${\cal F}_j(\lambda_j)$ by replacing $\vec{u}\cdot\vec{s}_N\to 
(1/9)[1+2(m_e/E_e)]\vec{u}\cdot\vec{s}_e$ and multiplying the whole expression 
by 1/2. Therefore, the discussion of the summation over the neutrino 
helicities, non-relativistic and ultra-relativistic neutrino limits and the 
differences between the Dirac and Majorana neutrino detection rates in the 
unpolarized case is quite similar to that for neutrino capture on 
polarized targets in section~\ref{sec:pol} and will not be repeated here.

\section{\label{sec:signoise}Separation of periodic signal from large 
fluctuating background}

The amplitude of periodic variations of the angular correlations in relic 
neutrino captures on beta-decaying nuclei is expected to be only a small 
($\sim 0.1\%$) fraction of the signal itself. In addition, except in the case 
of very high energy resolution, the overall relic neutrino signal 
is going to be very small compared with the background 
coming from the usual $\beta$ decay of the target nuclei. Indeed, the ratio of 
neutrino capture rate to the rate of the competing $\beta$ decay with 
production of electrons in the energy window $\delta$ just below the endpoint 
of the $\beta$-spectrum is \cite{cocco1,vogel1} 
\be
\frac{\Gamma_c}{\Gamma_d}\simeq 6\pi^2 \frac{n_\nu}{\delta^3}\simeq 
\frac{n_\nu}{56\;\mbox{cm}^{-3}}\frac{2.54\times 10^{-11}}
{[\delta\,(\rm eV)]^3}\,,
\label{eq:ratio}
\ee
independently of the $Q_\beta$-value of the process. Thus, e.g., for 
$n_\nu=56$ cm$^{-3}$ and $\delta=0.1$ eV the ratio is about 
$2.5\times 10^{-8}$. Clearly, in looking for the time dependence of the relic 
neutrino signal one would face a very difficult task of extracting a weak 
periodic signal from large fluctuating backgrounds. 

The problem of separation of signals from noisy backgrounds is actually well 
studied. It is routinely encountered in astronomy, acoustics, 
radiodetection, engineering, medical applications such as electrocardiography, 
magnetic resonance imaging, etc. There exists an enormous literature on the 
subject. Here we shall consider a simple approach to 
signal-from-noise separation based on the Fourier analysis and filtering in 
the frequency domain.   
 
We are interested in observing small periodic variations of the  detection 
rate of electrons produced in neutrino capture on polarized nuclei in the 
presence of a large background of electrons coming from the usual $\beta$ decay 
and possibly from other background sources. We also include in our definition 
of the background a (small) contribution coming from the average neutrino 
capture rate. The periodic component of the signal is thus defined to have 
zero time average. The background will be considered to be time-independent, 
except for the usual statistical fluctuations due to the quantum nature of 
the underlying processes.%
\footnote{
For large overall observation time $T$ comparable to the mean lifetime of 
the target nuclei one should take into account the exponential decrease 
with time of both the amplitude of the periodic signal and 
the background. In this case the mean background to be subtracted from 
the signal in order to ensure that it has zero average 
should be represented by a time-dependent moving average.}

In extracting periodic signals from large backgrounds the main problem 
comes from the fluctuations of the background rather than from its average.  
We therefore subtract the mean rate of the background events from the overall 
signal and consider  
\be
f(t)=s(t)+n(t)\,,
\label{eq:sig1}
\ee
where $s(t)$ is the periodic signal and the $n(t)$ is the ``noise'' 
which represents statistical fluctuations of the background. 
This is actually the form usually considered in the signal processing theory. 
Since by their definition both the signal $s(t)$ and the noise $n(t)$ have 
zero time averages, their strengths are characterized by their variances, 
$\overline{s(t)^2}$ and $\overline{n^2(t)}$.  
 
In our treatment we will consider the periodic signal of the sinusoidal form 
\be
s(t)=A_0\sin(\omega_0 t+\varphi)\,
\label{eq:sig2}
\ee
with the period $T_0=2\pi/\omega_0$ and constant amplitude $A_0$ and phase 
$\varphi$.%
\footnote{
For neutrino capture on polarized nuclei, $A_0$ is only 
constant when the degree of polarization of the target nuclei does not change 
with time. Depolarization effects will lead to a decrease of $A_0$ with time, 
and re-polarization of the target nuclei in the course of the experiment 
(or between the runs) may be necessary. Possible time dependence of $A_0$ 
can be readily incorporated in the statistical treatment of the signal.} 
The statistical properties of signal and noise are usually described 
in the time domain by their autocorrelation functions, $R_{ss}(\tau)$ and 
$R_{nn}(\tau)$, and in the frequency domain by their power spectra 
\cite{sig1,sig2,sig3,sig4,lomb,scargle}. Let us first consider the limit of 
infinite total observation time $T$ and infinite measurement bandwidth. The 
autocorrelation function for the signal is defined as   
\be
R_{ss}(\tau)=\overline{s(t)s(t+\tau)}\equiv\lim_{T\to\infty}\frac{1}{T}
\int_{-T/2}^{T/2}s(t)s(t+\tau) dt\,, 
\label{eq:autocor1}
\ee
and similarly for the noise. The signal and noise power spectra $S(\omega)$ 
and $N(\omega)$ are the Fourier transforms of the corresponding 
autocorrelation functions:  
\be
S(\omega)=\int_{-\infty}^\infty d\tau\,e^{-i\omega \tau}R_{ss}(\tau)\,,\qquad
N(\omega)=\int_{-\infty}^\infty d\tau\,e^{-i\omega\tau}R_{nn}(\tau)\,. 
\label{eq:sigS1}
\ee
If the power spectra are known, the autocorrelation functions 
can be obtained as inverse Fourier transforms:  
\be
R_{ss}(\tau)=\int_{-\infty}^\infty\frac{d\omega}{2\pi}\,e^{i\omega \tau}
S(\omega)\,,\qquad
R_{nn}(\tau)=\int_{-\infty}^\infty\frac{d\omega}{2\pi}\,e^{i\omega \tau}
N(\omega)\,.
\label{eq:sigS2}
\ee
Note that the power spectrum of the signal can also be found as
\be
S(\omega)=\lim_{T\to\infty}\frac{1}{T}\left|\tilde{s}(\omega)\right|^2\,,
\label{eq:sigS}
\ee 
where $\tilde{s}(\omega)$ is the Fourier transform of the time-dependent 
signal $s(t)$:
\be
\tilde{s}(\omega)=\int_{-\infty}^\infty dt\, e^{-i\omega t}s(t)\,.
\label{eq:fourier1}
\vspace*{-1.5mm}
\ee
This can be readily shown by substituting $R_{ss}(\tau)$ from 
(\ref{eq:autocor1}) into the first equation in (\ref{eq:sigS1}).

The autocorrelation function of a periodic function of period $T_0$ 
is itself a periodic function of the time lag $\tau$ with 
the same period. In particular, for the sinusoidal periodic 
signal (\ref{eq:sig2}) we find 
\be
R_{ss}(\tau)=\frac{A_0^2}{2}\cos\omega_0\tau\,.
\label{eq:autocor2}
\ee
The power spectrum of such a signal consists of two discrete lines 
corresponding to the frequencies $\omega=\pm\omega_0$: 
\be
S(\omega)=\frac{\pi}{2}
A_0^2\big\{\delta(\omega-\omega_0)+\delta(\omega+\omega_0)\big\}\,. 
\label{eq:sigS3}
\ee
The autocorrelation functions of random time distributions 
take their maximum values at $\tau=0$ and
tend to zero as $\tau\to\infty$. 
In the case of interest to us, the noise $n(t)$ is due to 
the fluctuations of the background event rate. 
As the values of these fluctuations at different times are completely 
statistically independent, 
all the frequencies contribute to the power spectrum of the noise with the 
same weight, i.e. 
\be
N(\omega)=N_0=const.
\label{eq:flat}
\ee
Such random time distributions with flat power spectrum are called white 
noise. The corresponding autocorrelation function 
\be
R_{nn}(\tau)=N_0\delta(\tau) 
\label{eq:autocor3}
\ee
vanishes for all $\tau\ne 0$. 

Within this approach it would be formally possible to separate arbitrarily weak 
periodic signal from noise: it would be sufficient to consider their 
autocorrelation functions at any non-zero time lag $\tau$ satisfying 
$\omega_0\tau\ne \pi/2+\pi n$. The noise autocorrelation 
function~(\ref{eq:autocor3}) would then vanish, while the signal 
autocorrelation function $R_{ss}$ of eq.~(\ref{eq:autocor2}) would 
remain finite. This, however, would have only been the case for infinite $T$, 
whereas all experiments have finite duration. Another indication of the 
oversimplified nature of the above consideration is that the total noise power 
obtained by integrating the flat power spectrum of eq.~(\ref{eq:flat}) over 
the whole frequency interval $(-\infty,\,\infty)$ is infinite. 
The resolution comes from the observation that any realistic measurement 
actually has a finite bandwidth, i.e.\ is characterized by a finite interval 
of frequencies. We therefore turn now to the realistic case of finite total 
observation time $T$ and finite measurement bandwidth. 

In experiments with time-dependent signal, quite often time 
binning of the data is invoked, either because of the experimental conditions 
or basing on some statistical considerations. 
For the bin length $\Delta t$, the signal will then represent a finite 
discrete time series, similar to those 
obtained by sampling continuous signals with the sampling rate 
$1/\Delta t$. The bin length, however, cannot be too large: the lossless 
reconstruction of a signal is in general only possible if the sampling rate 
exceeds twice the maximal linear frequency contained in the signal 
\cite{sig1,sig2,sig3,sig4,lomb,scargle}. For a simple periodic 
signal of period $T_0$ this means that the sampling interval 
$\Delta t$ must be below $T_0/2$. 

For a time series with a uniform sampling interval $\Delta t$ the 
bandwidth is limited by $|\omega|\le \omega_m$, where the maximal frequency   
$\omega_m$ is related to the Nyquist linear frequency 
$f_{\rm N}\equiv (2\Delta t)^{-1}$ by 
\be
\omega_m=2\pi f_{\rm N}=\frac{\pi}{\Delta t}\,.
\label{eq:Nq1}
\ee
This is essentially the highest frequency about which there is information, 
because $\Delta t$ is the shortest time interval spanned.  
In the case of uneven sampling, when the lengths of the sampling intervals 
vary, a generalized Nyquist linear frequency can be introduced, defined 
as $(2\Delta t)^{-1}$  with $\Delta t$ being the mean sampling interval 
\cite{scargle}. In experiments allowing real-time detection of the signal, 
binning of the data is not necessary and may actually be detrimental to 
detection of time-dependent signals \cite{lisi}, as any binning leads to 
information loss. In such real-time detection experiments $\Delta t$ can be 
taken to be the mean time interval between two consecutive events, i.e.\ 
the reciprocal of the mean event rate. 

Let us now consider the signal-to-noise ratio and its improvement by 
frequency-domain filtering, taking into account that the total 
observation time $T$ is finite. In what follows we will be assuming 
$\Delta t\ll T$ and for simplicity will replace the summation 
over the event detection times by integration, which corresponds to the limit 
$\Delta t\to 0$. The finite actual length of $\Delta t$ will only 
be reflected in that the integration in the frequency domain will be limited by 
the interval $[-\omega_m,\,\omega_m]$. More refined approach 
would be to estimate the signal power spectrum  by calculating periodograms 
based on discrete Fourier transform \cite{lomb,scargle}; however for our 
purpose of obtaining simple estimates the approach we adopt is quite 
adequate. 

\begin{figure}[h]
\centerline{\includegraphics[width=12cm,height=5.5cm]{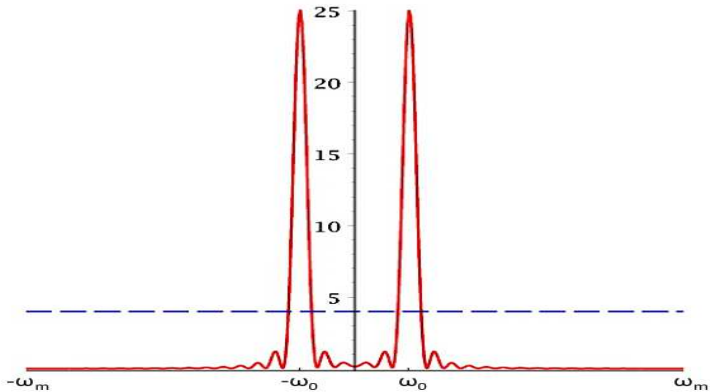}}
\caption[]{
Schematic representation of the power spectra of the 
signal $S(\omega)$ (red solid curve) and noise $N(\omega)$ 
(blue dashed line). 
}
\label{fig:1}
\end{figure}

Let us first calculate the signal power spectrum. 
It can be shown that $S(\omega)$ has the same form as that in 
eq.~(\ref{eq:sigS}) except that 
no limit $T\to\infty$ is taken and $\tilde{s}(\omega)$ is now the 
finite-interval Fourier transform of $s(t)$. Direct calculation with $s(t)$ 
from eq.~(\ref{eq:sig2}) yields 
\be
\tilde{s}(\omega)=\int_{-T/2}^{T/2} dt\,e^{-i\omega t}s(t)
=A_0\Big\{-ie^{i\varphi}\,
\frac{\sin\big[(\omega-\omega_0)\frac{T}{2}\big]}{\omega-\omega_0}
+ie^{-i\varphi}\,
\frac{\sin\big[(\omega+\omega_0)\frac{T}{2}\big]}{\omega+\omega_0}
\Big\}.
\label{eq:sigS2a}
\ee
The power spectrum of the signal is then 
\be
S(\omega)=\frac{A_0^2}{T}\Big\{
\frac{\sin^2\big[(\omega-\omega_0)\frac{T}{2}\big]}{(\omega-\omega_0)^2}
+\frac{\sin^2\big[(\omega+\omega_0)\frac{T}{2}\big]}{(\omega+\omega_0)^2}
-2
\frac{\sin\big[(\omega-\omega_0)\frac{T}{2}\big]}{\omega-\omega_0}
\frac{\sin\big[(\omega+\omega_0)\frac{T}{2}\big]}{\omega+\omega_0}\cos2\varphi
\Big\}.
\label{eq:sigS3a}
\ee
It contains two main peaks with the centers at 
$\omega=\pm \omega_0$ (see fig.~\ref{fig:1}). The widths of each peak can be 
characterized by the distance between the two zeros nearest to the peak 
position; this corresponds to the intervals $\pm 2\pi/T$ around the centers 
of the peaks.
Since the distance between the centers of the two main peaks $2\omega_0$ 
is much larger than the width of the peaks $4\pi/T$,%
\footnote{We assume that the total observation time $T$ is much larger than 
the period of the sidereal daily variations of the signal $2\pi/\omega_0
\simeq 24$\,h.}
the peaks are practically non-overlapping, and one can safely neglect the 
cross-product term (the last term in the curly brackets) in 
eq.~(\ref{eq:sigS3a}). The interval $[-2\pi/T,\, 2\pi/T]$ around the center 
of each peak contains about 90\% of its strength (see eq.~(\ref{eq:sig3}) 
below). In the limit $T\to \infty$ we recover the result in 
eq.~(\ref{eq:sigS3}). 

It was pointed out above that the signal and noise strengths can be 
characterized by their variances, $\overline{s(t)^2}$ and $\overline{n^2(t)}$. 
As follows from~(\ref{eq:autocor1}) and the similar definition of 
$R_{nn}(\tau)$, these variances coincide with the values of the corresponding 
autocorrelation functions at the origin, which in turn are given by the 
integrals of the signal and noise powers over the bandwidth $\Omega$ of the 
measurement:
\be
\overline{s(t)^2}=R_{ss}(0)=\int_\Omega\frac{d\omega}{2\pi}S(\omega)\,,\qquad
\overline{n(t)^2}=R_{nn}(0)=\int_\Omega\frac{d\omega}{2\pi}N(\omega)\,.
\label{eq:strengths}
\ee  
As the power spectrum of the background fluctuations 
$N(\omega)$ is essentially constant throughout the bandwidth of 
the experiment $[-\omega_m,\,\omega_m]$, we have 
\be
\overline{n(t)^2}\equiv\sigma_0^2=\int_{-\omega_m}^{\omega_m}
\frac{d\omega}{2\pi}\,N(\omega)\simeq \frac{\omega_m}{\pi}N_0\,.
\label{eq:sigN2}
\ee
In the case under consideration, the noise is due to the statistical  
fluctuations of the background related to quantum nature of the underlying 
processes and discrete nature of the detected particles; this is a 
special case of white noise called shot noise \cite{sig4}. The flat power 
spectrum of such a noise is given by the number of events per unit time 
(see e.g.\ \cite{chen} for a simple derivation). Thus, the quantity $N_0$ in 
eq.~(\ref{eq:sigN2}) is just the mean background event rate. 

We will now make use of the fact that the signal is concentrated in the 
narrow intervals in the frequency domain (which become narrower as the 
total observation time increases), whereas the noise power is constant 
per unit bandwidth. To improve the signal-to-noise ratio, we therefore 
integrate the signal and noise power spectra over the intervals 
$\Delta\omega=[-2\pi/T,\,2\pi/T]$ around the points $\omega=\pm\omega_0$. 
Since the regions around these two points 
give identical contributions to the signal power as well as to the noise 
power, for the ratio it is sufficient to consider only one such region. 
We therefore define the improved signal-to-noise ratio as 
\be
\rho=\frac{\int_{\Delta\omega}\frac{d\omega}{2\pi}S(\omega)}
{\int_{\Delta\omega}\frac{d\omega}{2\pi}N(\omega)}\,.
\label{eq:rho1}
\ee
With the expression for $S(\omega)$ from eq.~(\ref{eq:sigS3a}) 
and $N(\omega)=N_0$ we find 
\be
\int_{\Delta\omega}\frac{d\omega}{2\pi}S(\omega) 
\simeq \frac{A_0^2}{4\pi}\int_{-\pi}^\pi \frac{\sin^2 x}{x^2} dx
=\frac{A_0^2}{4}\,\frac{2{\rm Si}(2\pi)}{\pi}
\simeq\frac{A_0^2}{4}\cdot 0.9028\,,
\label{eq:sig3}
\ee
\be
\int_{\Delta\omega}\frac{d\omega}{2\pi}N(\omega)\simeq \frac{2N_0}{T}\,. 
\label{eq:sigN4}
\ee
Substituting this into eq.~(\ref{eq:rho1}) yields 
\be
\rho\simeq \frac{0.9028\, A_0^2 T}{8N_0}\,.
\label{eq:rho2}
\ee 
This quantity essentially coincides (up to about a factor of 1/2) with the 
ratio of the peak value of the signal power to the (flat) value of the noise 
power. Comparing this result with the original signal-to-noise ratio 
$\overline{s(t)^2}/\overline{n(t)^2}$ that may be found 
by integrating the signal and noise power spectra over the full interval 
$[-\omega_m,\,\omega_m]$, we find that the improvement factor due to the 
passband filtering is about $0.225 N_0 T$. As $N_0 T$ is the number 
of background events detected over the total observation time $T$, this 
factor can be quite large. 

An alternative interpretation of the obtained result can be given by comparing 
it with the signal-to-background ratio 
rather than with the ``un-filtered'' signal-to-noise one.  
To this end, let us rewrite eq.~(\ref{eq:rho2}) as 
\be
\rho\simeq 
\frac{(A_0/\sqrt{2})T}{N_0 T}\cdot \left[0.225 (A_0 T/\sqrt{2})\right]. 
\label{eq:rho3}
\ee 
Here the quantity 
$(A_0/\sqrt{2})T$ is the root mean square number of the ``signal events'' 
(i.e.\ of the events caused by the periodic component of the neutrino capture 
signal) over the time interval $T$. The first factor in eq.~(\ref{eq:rho3}) is 
therefore essentially the signal-to-background ratio. The expression 
in the square brackets is the enhancement factor, which increases with 
increasing signal statistics $(A_0/\sqrt{2})T$. 

{}From eq.~(\ref{eq:rho3}) it can be seen that, though the noise reduction by 
simple frequency-domain filtering considered here will greatly improve the 
signal-to-noise ratio, it will not be sufficient for a reliable signal 
detection unless the total number of the detected signal events is very large. 
Indeed, for $\delta\sim 0.1$ eV and $u\sim 10^{-3}$ the signal-to-background 
ratio is expected to be of order $2.5\times 10^{-11}$. Thus,  with the 
simple frequency-domain filtering discussed above, an unrealistically large 
number of signal events ($\gtrsim 2.5\times 10^{11}$) would be necessary in 
order to detect the C$\nu$B signal. 
Therefore, more sophisticated methods of noise reduction must be invoked. 
As an example, the signal-to-noise ratio can be further improved by measuring 
the noise power spectrum in the frequency regions where the signal is 
essentially absent (i.e.\ away from the neighborhoods of 
$\omega=\pm\omega_0$) in order to predict its value in the regions where the 
signal is concentrated and cancel (subtract) it there \cite{samsing}. The 
cancellation will, however, be incomplete as the noise spectrum is in reality 
only approximately flat. 
Preliminary estimates show that such noise cancellation may 
reduce the requisite number of signal events to about $10^4$. Other methods 
of noise suppression can also be used, such as e.g.\ studying 
cross-correlation of the observed data with test functions in the form of 
the expected time-dependent signal,
as it is currently done in gravitational wave detection. 

To be able to make a more definitive statement on the 
possibility of noise suppression, one would need to perform signal and noise 
simulations in the conditions of a particular experiment.

\section{\label{sec:disc}Discussion}

We have suggested an alternative method of detection of C$\nu$B through 
neutrino capture on beta-decaying nuclei, based on observation of time 
variations of the total detection rates or of various angular correlations 
characteristic of $\beta$-processes. The time dependence should come about 
because of the peculiar motion of the Sun with respect to C$\nu$B and rotation 
of the Earth about its axis. The variations of total C$\nu$B detection rate 
can be observed in experiments with polarized nuclear targets, whereas 
the observation of modulated $\beta-\nu$ angular correlation or of the 
correlation between the neutrino direction and the spin of the produced 
electron does not require target polarization. 
As an example, we considered neutrino detection through allowed Gamow-Teller 
$1^\pi\to 0^\pi$ $\beta$-transitions; however, our results can be readily 
generalized to arbitrary allowed Gamow-Teller transitions 
$J^\pi\to J^\pi\pm 1$. 

One advantage of pure Gamow-Teller transitions is that the 
effects of the target polarization are more prominent in this case. 
Consider e.g. pure Gamow-Teller $J^\pi\to J^\pi-1$ transitions 
(of which  the $1^\pi\to 0^\pi$ process we have studied is an example). 
In this case the detection cross section will be exactly zero if the 
incoming neutrino is in a helicity eigenstate and its spin is aligned 
with that of the parent nucleus. Indeed, in this case the total 
angular momentum of the initial state (incoming neutrino + parent nucleus) 
is $J+1/2$; due to angular momentum conservation,  
such a system cannot decay into a daughter nucleus with spin 
$J-1$ and electron. Likewise, the probability of the process 
vanishes if the produced electron is in a helicity eigenstate with its spin 
antialigned with the polarization of the parent nucleus. It is easy to see 
that our squared amplitude (\ref{eq:product2}) satisfies these conditions 
(see footnote~\ref{fn:note1}). The case of pure Gamow-Teller 
$J^\pi\to J^\pi+1$ transitions can be considered quite similarly. 
 
On the other hand, pure Fermi transitions $0^\pi\to 0^\pi$ may be useful for 
observing $\beta-\nu$ angular correlations in relic neutrino capture 
on unpolarized nuclei, as they are characterized by a sizeable correlation 
coefficient $a=1$. An important advantage of pure transitions is that 
for them the angular correlations coefficients do not depend on nuclear 
matrix elements. 

For the non-relativistic component of C$\nu$B, the coefficients of angular 
correlations in relic neutrino capture 
are different for Dirac and Majorana neutrinos, and so one might hope that 
observing these correlations would help establish neutrino nature. 
Unfortunately, this would be extremely difficult, as the differences between 
the correlation coefficients for Dirac and Majorana neutrinos are additionally 
suppressed by the small factor $v_j$.

The suggested method of C$\nu$B detection is obviously prone to difficulties. 
One of them is related to the expected smallness of the amplitude of the 
periodic variation of the relic neutrino signal.  It will therefore be 
necessary to extract a weak periodic signal from a large  
fluctuating background of electrons or positrons from the usual $\beta$-decay. 
In experiments with polarized nuclear targets one will also have to face 
experimental problems related with the requirement of polarization of the 
target nuclei and maintaining this polarization for an extended period of time 
(or renewing it in the course of the experiment). 

On the other hand, there are well-developed powerful methods of 
signal-from-noise separation, which are especially efficient when the signal 
is of known periodicity and which therefore may be helpful in solving the 
problem of detecting relic neutrinos through time variations of their signal. 
The C$\nu$B detection may also be facilitated by the fact that there exist 
several independent angular correlations which are expected to exhibit time 
variations of the same periodicity. It is also worth noticing that the 
the angular correlations in the relic neutrino capture may be enhanced in the 
presence of non-standard neutrino physics, such as relatively large neutrino 
magnetic moments \cite{lisanti1}. 

The main advantage of the approach suggested in this paper is that it does 
not in principle require extremely high energy resolution of the detector. 
Good energy resolution would certainly be helpful, as it would allow one to 
suppress the background of electrons or positrons coming from the competing 
$\beta$ decay by working close to the endpoint of their spectrum. However, 
the method can still be operative even for relatively large energy resolution, 
if sufficiently potent method of separation of the weak periodic signal from 
strong random noise is employed. This is in contrast with the situation 
with the usually considered method of relic neutrino detection based on 
the separation of the spectra of $\beta$-particles from neutrino absorption 
and $\beta$ decay. The latter would not work if the energy resolution of the 
detector exceeds the largest neutrino mass.  
The fact that the requirements on the energy resolution are less severe in 
the approach suggested here, in particular, means that radioactive 
nuclei with relatively large  $Q_\beta$ values may be preferable as target, 
because they lead to larger absolute detection rates. 

Much work has yet to be done to establish if the approach to C$\nu$B detection 
suggested in this paper is actually feasible, including selection of suitable 
nuclides (taking into account, among other aspects, their availability, 
lifetime, type of $\beta$ transition, (neutrino capture)/($\beta$ decay) rate 
ratio and feasibility of target polarization in the case of experiments with 
polarized nuclei). Extensive simulations  of the signal and background will 
also be necessary to clarify if the proposed C$\nu$B detection method is 
actually viable and is competitive with the approach based on the separation 
of the spectra of $\beta$-particles from relic neutrino capture from those 
produced in $\beta$ decay of target nuclei.

\acknowledgments
The author is grateful to Alexei Smirnov for numerous useful discussions 
and to Eligio Lisi for useful correspondence.

\appendix
\section{Angular averaging}

We will consider here the effects of averaging over the angular distributions 
of relic neutrinos on the angular correlations of interest in the lab frame. 
We will be assuming that in the C$\nu$B rest frame neutrinos have isotropic 
velocity distribution and are in the exact helicity eigenstates; therefore 
their spins are also distributed isotropically. 
As in section~\ref{sec:movframe}, we will use the unprimed notation for the 
neutrino variables in the C$\nu$B rest frame, whereas the primed quantities 
will refer to the lab frame, moving with the velocity $-\vec{u}$ with respect 
to the cosmic one. We will consider the effects of Lorentz boost from the 
C$\nu$B rest frame to the lab frame to first order in $u$. The neutrino 
energy, momentum and velocity in the lab frame are given in eq.~(\ref{eq:Ep}). 
The expressions for neutrino spin in the lab frame are given in 
eq.~(\ref{eq:s}) in the general case and in eq.~(\ref{eq:s2}) for neutrinos that 
are in helicity eigenstates in the C$\nu$B rest frame. It was demonstrated in 
section~\ref{sec:movframe} that  
in the case $v_j\gg u$ of main interest to us neutrinos that were in states 
of definite helicity in the C$\nu$B frame will remain in the same helicity 
states in the lab frame up to corrections ${\cal O}(u^2)$ which we neglect. 

{}From eqs.~(\ref{eq:Ep}) and (\ref{eq:s2}) we find 
\be
\vec{v}\,'\!\!_j\!\cdot\!\vec{s}\,'\!\!_j=\lambda_j v_j\Big\{1+
(\vec{u}\!\cdot\!\vec{v}_j)\frac{1-v_j^2}{v_j^2}\Big\}\,. 
\label{eq:s2aa}
\ee
For the quantity $K_j$ that enters into the expression for the neutrino 
vector $B^\mu$ we then obtain in the lab frame 
\be
K_j'=1-\frac{E'\!\!_j}{E'\!\!_j+m_j}(\vec{v}\,'\!\!_j\!\cdot\!\vec{s}\,'\!\!_j)=
1-\frac{E_j}{E_j+m_j}\lambda_j v_j \Big(1+\frac{\vec{u}\!\cdot\!\vec{v}_j}
{v_j^2}\frac{m_j}{E_j}\Big)\,.
\label{eq:Kj2aa}
\vspace*{2.5mm}
\ee
This gives 
\be
\frac{1}{E_j'}\vec{B}'
\,=\,K_j'\vec{v}\,'\!\!_j-\frac{m_j}{E_j'}\vec{s}\,'\!\!_j
\,=\,(1-\lambda_j v_j)\Big\{\vec{u}-
(1-\vec{u}\!\cdot\!\vec{v}_j)\lambda_j\frac{\vec{v}_j}{v_j}\Big\}
\,.
\label{eq:Kj3aa}
\ee

Consider now the effects of averaging over the angular distributions of 
relic neutrinos on neutrino variables, taking into account that neutrino 
velocity and spin distributions in the C$\nu$B rest frame are isotropic. 
Denoting the averaging over the directions by angular brackets and taking 
into account that $\langle\vec{v}_j\rangle=0$ and $\langle\vec{s}_j\rangle=0$, 
for the quantities in the lab frame we find 
\be
\langle E_j'\rangle=E_j\,,\qquad \langle\vec{q}\,'\rangle=
\vec{u} E_j\,,\qquad \langle \vec{v}\,'\!\!_j\rangle=\Big(1-\frac{v_j^2}{3}\Big)
\vec{u}\,,
\label{eq:Epaveraa}
\ee
\be
\Big\langle\frac{\vec{v}\,'\!\!_j}{v_j'}\Big\rangle=\frac{2}{3}\frac{\vec{u}}
{v_j}\,,
\vspace*{1.5mm}
\label{eq:vvaveraa}
\ee
\be
\langle\vec{s}\,'\!\!_j\rangle=\lambda_j v_j\,\frac{2}{3}\frac{E_j}{E_j+m_j}
\vec{u}\,,
\qquad
\big\langle\vec{v}\,'\!\!_j\!\cdot\!\vec{s}\,'\!\!_j\big\rangle
=\lambda_j v_j=\big\langle\vec{v}_j\!\cdot\!\vec{s}_j\big\rangle\,.
\label{eq:S2averaa}
\vspace*{1.5mm}
\ee
This \vspace*{2.5mm}yields  
\be
\Big\langle\frac{1}{E_j'}B^{0}{}'
\Big\rangle=1-\langle
\vec{v}\,'\!\!_j\!\cdot\!\vec{s}\,'\!\!_j \rangle=1-\lambda_j v_j\,,
\qquad\qquad\qquad~
\label{eq:B0Aver} 
\ee
\be
\Big\langle \frac{1}{E_j'}\vec{B}'
\Big\rangle
=
\Big\langle K_j'\vec{v}\,'\!\!_j-\frac{m_j}{E_j'}\vec{s}\,'\!\!_j\Big\rangle=
\Big(1-\frac{2}{3}\lambda_j v_j-\frac{v_j^2}{3}\Big)\vec{u}\,.
\label{eq:BvecAver}
\ee
With these relations, one readily obtains the expressions for the squared 
amplitude of the process given in sections~\ref{sec:pol} and~\ref{sec:unpol}. 

\section{ 
Electron asymmetry with respect to a fixed direction in the lab frame}

As discussed in section~\ref{sec:unpol}, the angular correlation between the 
direction of the produced electron and the preferred direction of the neutrino 
arrival in the case of relic neutrino capture on unpolarized nuclei can be 
written as 
\be
const.(1+\alpha \vec{u}\!\cdot\!\vec{v_e})\,,
\label{eq:corr1}
\ee
where $\alpha$ is the correlation coefficient (see eq.~(\ref{eq:factor3a})).  
We want to find the time-dependent forward-backward asymmetry of electron 
emission 
with respect to a fixed direction in the lab frame 
specified by a unit vector $\vec{\xi}$. It is easy to see that the asymmetry 
would not depend on time for $\vec{\xi}$ collinear with the rotation axis 
of the Earth and its time dependence will be maximized for orthogonal 
directions. We therefore choose $\vec{\xi}$ to lie in the plane orthogonal 
to the Earth's rotation axis. In the geocentric spherical coordinates we 
have 
\begin{align}
&\vec{u}=u\big(\cos\phi_u(t)\sin\theta_u\,,\sin\phi_u(t)\sin\theta_u\,, 
\cos\theta_u\big)\,,
\label{eq:u}
\\
&\vec{v}_e=v_e\big(\cos\phi_e\sin\theta_e\,,\sin\phi_e\sin\theta_e\,, 
\cos\theta_e\big)\,,
\label{eq:ve}
\\
&\vec{\xi}=\big(\cos\phi_\xi\,,\sin\phi_\xi\,, 0\big)\,,
\label{eq:xi}
\end{align}
where 
\be
\phi_u(t)=\frac{2\pi}{T_0}t+\phi_0
\ee
and $T_0\simeq 24$\,h is the sidereal day. From \ref{eq:u} and \ref{eq:ve} 
we find 
\be
\vec{u}\!\cdot\!\vec{v}_e=u v_e
\big[\cos\theta_u\cos\theta_e+
\sin\theta_u\sin\theta_e\cos[\phi_u(t)-\phi_e]\big]\,.
\label{eq:}
\ee
Straightforward calculation then gives for the forward-backward asymmetry 
of the electron emission with respect to $\vec{\xi}$
\be
{\cal A}\equiv 
\frac{\sigma_\uparrow-\sigma_\downarrow}
{\frac{1}{2}[\sigma_\uparrow+\sigma_\downarrow]}=\alpha u v_e \sin\theta_u 
\cos[\phi_u(t)-\phi_\xi]\,.
\ee
The electron spin asymmetry with respect to a fixed direction in the 
lab frame can be considered quite similarly.

\end{document}